\documentclass[12pt]{article}
\usepackage{amsmath}
\usepackage{amsthm}
\usepackage{amsfonts}
\usepackage{amssymb}

\input epsf

\newtheorem{thm}{Theorem}[section]

\newtheorem{lemm}[thm]{Lemma}

\newcommand{\C}{{\mathbb C}}
\newcommand{\Q}{{\mathbb Q}}

\def\U{{\bf U}}

\begin{document}

\title{A modular functor which is universal for quantum computation}
\author{Michael Freedman$^{\dag}$, Michael Larsen$^{\ddag}$,
and Zhenghan Wang$^{\ddag}$}

\maketitle $\dag$ {\it Microsoft Research, One Microsoft Way,
michaelf@microsoft.com}

$\ddag$ {\it Indiana Univ., larsen@math.indiana.edu and
zhewang@indiana.edu}

\begin{abstract}
We show that the topological modular functor from
Witten-Chern-Simons theory is universal for quantum computation in
the sense a quantum circuit computation can be efficiently
approximated by an intertwining action of a braid on the functor's
state space. A computational model based on Chern-Simons theory at
a fifth root of unity is defined and shown to be polynomially
equivalent to the quantum circuit model.  The chief technical
advance: the density of the irreducible sectors of the Jones
representation, have topological implications which will be
considered elsewhere.
\end{abstract}

\newpage

\section{Introduction}
The quantum computer was Feynman's [Fey] last great idea. He
understood that local $\lq\lq$quantum gates", the basis of his
model, can efficiently simulate the evolution of any finite
dimensional quantum system and by extension any renormalizable
system.  The details of the argument are given in [Ll].
Topological quantum field theories (TQFTs), although possessing a
finite dimensional Hilbert space, lack a Hamiltonian---the
derivative of time evolution on which the Feynman-Lloyd argument
is based. In [FKW], we provide a different argument for the
poly-local nature of TQFTs showing that quantum computers
efficiently simulate these as well.  Here we give a converse to
this simulation result.  The Feynman-Lloyd argument is reversible,
so we may summarize the situation as:

(1) finite dimensional quantum systems,

(2) quantum computers (meaning the quantum circuit model QCM
[D][Y]),

(3) certain topological modular functors (TMFs).

Each can efficiently simulate the others.  We wrote TMF above
instead of TQFT because we use only the conformal blocks and the
action of the mapping class groups on these---not the general
morphisms associated to 3-dimensional non-product bordisms.

We would like to thank Alexei Kitaev for conversations on our
approach.

\section{A universal quantum computer}

The strictly 2-dimensional part of a TQFT is called a {\it
topological modular functor} (TMF). The most interesting examples
of TMFs are given by the SU(2) Witten-Chern-Simons theory at roots
of unity [Wi].  These examples are mathematically constructed in
[RT] using quantum groups (See also [T][Wa]). A modular functor
assigns to a compact surface $\Sigma$ (with some additional
structures detailed below) a complex vector space $V(\Sigma)$ and
to a diffeomorphism of the surface (preserving structures) a
linear map of $V(\Sigma)$.  In the cases considered here
$V(\Sigma)$ always has a positive definite Hermitian inner product
$<,>_{h}$ and the induced linear maps preserve $<,>_{h}$, i.e. are
unitary.  The usual additional structures are fixed
parameterizations of each boundary component, a labeling of each
boundary component by an element of a finite label set
$\mathcal{L}$ with an involution $\hat{} : \mathcal{L} \rightarrow
\mathcal{L}$, and a Lagrangian subspace $L$ of $H_{1}(\Sigma,\Q)$
([T][Wa]). Since our quantum computer is built from
quantum-$SU(2)$-invariants of braiding, and the intersection
pairing of a planar surface is $0$, $L=H_{1}(\Sigma;\Q)$ and can
be ignored.  The parameterization of boundary components can be
dropped. (The essential information which enhances the Kauffman
bracket to the Jones polynomial is remembered by the
$\lq\lq$blackboard framing" of the braid.)  The involution
$\hat{}$  is simply the identity since the $SU(2)$-theory is
self-dual. In fact, we can manage by only considering the
$SU(2)$-Chern-Simons theory at $q=e^{\frac{2\pi i}{r}},r=5$ and so
our label set will be the symbols $\{0,1,2,3\}$.  Note that in our
notation, $0$ labels the trivial representation, not $1$. Since we
are suppressing boundary parameterizations, we may work in the
disk with $n$ marked points-thought of crushed boundary
components. Because we only need the $\lq\lq$uncolored theory" to
make a universal model, each marked point is assigned the label
$1$, and the boundary of the disk is assigned the label $0$.  We
consider the action of the braid group $B(n)$ which consists of
diffeomorphisms of the disk which leave the $n$ marked points and
the boundary set-wise invariant modulo those isotopic to the
identity. The braid group has the well-known presentation: $$
\begin{array}{cc}
B(n)=\{ \sigma_1, \cdots, \sigma_{n-1}| &
\sigma_{i}\sigma_{j}\sigma_{i}^{-1} \sigma_{j}^{-1}=id\;\;
\text{if} \;\; |i-j|>1  \\
 & \sigma_{i}\sigma_{j}\sigma_{i}=\sigma_{j}\sigma_{i}\sigma_{j}
\;\; \text{if}\;\; |i-j|=1\},
\end{array} $$
where $\sigma_{i}$ is the half right twist of the $i$-th marked
point about the $i+1$-st marked point.

To describe our fault-tolerant computational model
$\lq\lq$Chern-Simons5" {\bf CS5}, we must deal with the usual
error arising from decoherence as well as a novel $\lq\lq$qubit
smearing error" resulting from imbedding the computational qubits
within a modular functor super-space.  To explain our approaches
we initially ignore all errors; in particular formula (\ref{2.0})
is a simplification valid only in the error-free context.

The state space $S_{k}=(\C^{2})^{\otimes k}$ of our quantum
computer consists of $k$ qubits, that is the disjoint union of $k$
spin=$\frac{1}{2}$ systems which can be described mathematically
as the tensor product of $k$ copies of the state space $\C^{2}$ of
the basic 2-level system, $\C^{2}=\text{span}(|0>, |1>)$.  For
each even integer $k$, we will choose an inclusion $
S_{k}\stackrel{i}{\hookrightarrow} V(D^{2}, \;\; \text{3k \;
marked\; points}) =V(D^{2}, 3k)$ and show how to use the action of
the braid group $B(3k)$ on the modular functor to (approximately)
induce the action of any poly-local unitary operator $\U:
S_{k}\rightarrow S_{k}$.  That is we will give an (in principle)
efficient procedure for constructing a braid $b=b(\U)$ so that
\begin{equation}
 i\circ \U=V(b)\circ i. \label{2.0}
\end{equation}
To see that this allows us to simulate the QCM, we need to
explain: $(i)$ what we mean by the hypothesis $\lq\lq$poly-local"
on $\U$, $(ii)$ what $\lq\lq$efficient" means, $(iii)$ what the
effect of the two types of errors are on line (\ref{2.0}), and
$(iv)$ what measurement consists of within our model.

We begin by explaining how to map $S_{k}$ into $V$ and how to
perform 1 and 2 qubit gates.

Let $D$ be the unit 2-dimensional disk and
$$
\left\{
\frac{11}{100k}, \frac{12}{100k}, \frac{13}{100k},
\frac{21}{100k}, \frac{22}{100k}, \frac{23}{100k}, \cdots,
\frac{10k+1}{100k}, \frac{10k+2}{100k}, \frac{10k+3}{100k}
\right\}$$
be a subset of $3k$ marked points on the $x$-axis.  Without giving
formulae the reader should picture $k$ disjoint sub-disks $D_{i},
1\leq i\leq k$, each containing one clump of $3$ marked points in
its interior (these will serve as qubits) and further $\left(
\begin{array}{c} k\\ 2
\end{array} \right)$
 disks $D_{i,j}, 1\leq i<j\leq k$, containing
$D_{i}$ and $D_{j}$, but with $D_{ij}\cap D_{l}=\emptyset, l\neq i
\;\text{or}\; j$ (which will allow 2-qubit gates).  Strictly
speaking, among the larger subdisks, we only need to consider
$D_{i,i+1}, 1\leq i, i+1<k$, and could choose a standard (linear)
arrangement for these but there is no cost in the exposition to
considering all $D_{i,j}$ above which will correspond in the model
to letting any two qubits interact.  Also, curiously, we will see
that any of the numerous topologically distinct arrangements for
the $\{D_{i,j}\}$ within $D$ may be selected without prejudice.

We define $V_{k}^{l}$ to be the $SU(2)$ Hilbert space of $k$
marked points in the interior with labels equal 1 and $l$ label on
$\partial D$.
  We need to understand the many ways in which $V^{0}_{m}$ arises
via the $\lq\lq$gluing axiom" ([Wa]) from smaller pieces.  The
axiom provides an isomorphism:
\begin{equation}
V(X\cup_{\gamma}Y)=\oplus_{\text{all consistent labelings}\;\; l}
V(X,l)\otimes V(Y,l), \label{2.1}
\end{equation}
where the notation suppress all labels not on the 1-manifold
$\gamma$ along which $X$ and $Y$ are glued.  The sum is over all
labelings of the components of $\gamma$ satisfying the conditions
that matched components have equal labels.  According to
$SU(2)$-Chern-Simons theory [KL],  for three-punctured spheres
with boundary labels $a,b,c$, the Hilbert space $V_{abc}\cong \C$
if

$$(i): a+b+c=\text{even},$$
\begin{equation}
(ii):a\leq b+c, b\leq a+b, c\leq a+b \;\;\text{(triangle
inequalities)} \label{2.2}
\end{equation}
$$(iii): a+b+c\leq 2(r-2);$$ and $V_{abc}=0$ otherwise.  The
gluing axiom together with the above information allows an
inductive calculation of $V_{k}^{l}$, where the superscript
denotes the label on $\partial D$.  We easily  calculate that
\begin{equation}
\text{dim}V_{3}^{1}=2, \;\; \text{dim}V_{3}^{3}=1,
\text{dim}V_{6}^{0}=5, \;\; \text{dim}V_{6}^{2}=8. \label{2.4}
\end{equation}

Line (\ref{2.4}) motivates taking $V(D_{i}, \text{its 3 pts},
\text{boundary label 1})$ $=:V_{i}\cong \C^{2}$ as our fundamental
unit of computation, {\it the qubit}.  We fix the choice of an
arbitrary $\lq\lq$complementary vector" $v$ in the state space of
$D\backslash \cup_{i=1}^{k}D_{i}$
\newline $v\in V(D\backslash \cup_{i=1}^{k}D_{i}$, all
boundary labels $= 1$ except boundary of  $D = 0$) \break
$=:V_{\text{complement}}$ (To keep this space nontrivial, we have
taken k even.)  Using $v$, the gluing axiom defines an injection:
\begin{equation}
i_{v}: ({\C}^{2})^{\otimes
k}\cong\otimes_{i=1}^{k}V_{i}\stackrel{\otimes v}{\rightarrow}
(\otimes_{i=1}^{k}V_{i})\otimes V_{\text{complement}}
\stackrel{\text{as summand}}{\hookrightarrow} V_{3k}^{0}
\label{2.6}
\end{equation}
This composition $i_{v}$ determines what we will serve as our
computational qubits within the modular functor $V_{3k}^{0}$.  The
reader familiar with [FKW] will notice that we use here a dual
approach.  In that paper, we imbedded the modular functor into a
larger Hilbert space that is a tensor power; here we imbedded a
tensor power into the modular functor.

The action of $B(3)$ on $D_{i}$ yields 1-qubit gates, whereas two
qubit gates will be constructed using the action of $B(6)$ on
$D_{i,j}$. Supposing our quantum computer $S_{k}$ is in state $s$,
a given $v$ as above determines a state $i_{v}(s)=s\otimes v\in
V_{3k}^{0}$.  Now suppose we wish to evolve $s$ by a 2-qubit gate
$g$ acting unitarily on $\C_{i}^{2}\otimes \C_{j}^{2}$ and by $id$
on $\C_{l}^{2}, l\neq i\; \text{or}\; j$.  Using gluing axiom
(\ref{2.1}) and the inclusion (5), we may write
\begin{equation}
s=\sum_{h}t_{h}\otimes u_{h}, \label{2.8}
\end{equation}
where $\{t_{h}\}$ is a basis or partial basis for
$\C^{2}_{i}\otimes \C^{2}_{j}$ and $u_{h}\in \otimes_{l\neq
i,j}\C_{l}^{2}$, so $s\otimes v=\sum_{h}(t_{h}\otimes
u_{h})\otimes v$. Decomposing along $\gamma=\partial D_{i,j}$, we
may write $v=\alpha_{0}\otimes \beta_{0}+\alpha_{2}\otimes
\beta_{2}$, where $\alpha_{\epsilon} \in V \big( D_{i,j}\backslash
(D_{i}\cup D_{j} ), \epsilon \;\text{on}\;\gamma \big)$, $\epsilon=0$ or
$2$ and $\beta_{\epsilon} \in V \big( D\backslash (\cup_{l\neq
i,j}D_{l}\cup D_{ij}), \epsilon \; \text{on} \;\gamma$, and $0$ on
$\partial D \big)$. Thus
\begin{equation}
s\otimes v=\sum_{h}t_{h}\otimes u_{h}\otimes \alpha_{0}\otimes
\beta_{0} +\sum_{h}t_{h}\otimes u_{h}\otimes \alpha_{2}\otimes
\beta_{2}, \label{2.10}
\end{equation}

An element of $B(6)$ applied to the 6 marked points in $D_{i}\cup
D_{j}\subset D_{ij}$ acts via a representation $\rho^{0}\oplus
\rho^{2}=:\rho$ on $V^{0}(D_{ij},\text{6 pts})\oplus V^{2}(D_{ij},
\text{6 pts}),$ where the superscript denotes the label appearing
when the surface is cut along $\gamma$.  In particular $B(6)$ acts
on each factor $t_{h}\otimes \alpha_{0}$ and $t_{h}\otimes
\alpha_{2}$ in (\ref{2.10}). Note $t_{h}\otimes \alpha_{0}$
belongs to the summand of $V^{0}(D_{ij}, \text{6 pts})$
corresponding to boundary labels on  $\partial \big(D_{ij}\backslash
(D_{i}\cup D_{j})\big)=0,1,1$. There is an additional 1-dimensional
summand corresponding to boundary labels 0,3,3-with 0,1,3 and
0,3,1 excluded by the triangle inequality $(ii)$ in (\ref{2.2})
above. Similarly $t_{h}\otimes \alpha_2$ belongs to the summand of
$V^{2}(D_{ij}, \text{6 pts})$ with boundary labels=2,1,1.  There
are additional summands corresponding to (2,1,3), and (2,3,1) of
dimensions 2 each.

Ideally we would find a braid $b=b(g)\in B(6)$ so that
$\rho^{0}(b)(t_{h}\otimes \alpha_{0})= gt_{h}\otimes\alpha_{0}$
and $\rho^{2}(b)(t_{h}\otimes \alpha_{2})=
gt_{h}\otimes\alpha_{2}$.
  Then
referring to (\ref{2.10}) we easily check that
\begin{equation}
\rho(b)(s\otimes v)=\sum_{h}\big( (gt_{h})\otimes
u_{h}\big)\otimes v, \label{2.12}
\end{equation}
i.e. $\rho(b)$ implements the gate $g$ on the state space $S_{k}$
of our quantum computer. In practice there are two issues: $(i)$
we cannot control the phase of the output of either $\rho^{0}$ or
$\rho^{2}$, and $(ii)$ these outputs will be only approximations
of the desired gate $g$.  The phase issue $(i)$ leads to a change
of the complimentary vector $v\rightarrow v'$ as follows as seen
on line $(\ref{2.9})$ below.  This is harmless since ultimately we
only measure the qubits.
 $$ s\otimes
v=\sum_{h}t_{h}\otimes u_{h}\otimes \alpha_{0}\otimes \beta_{0}
+\sum_{h}t_{h}\otimes u_{h}\otimes \alpha_{2}\otimes \beta_{2} $$
\begin{center}
$\Downarrow$ gate
\end{center}
$$ s\otimes v=\omega_{0} \sum_{h}gt_{h}\otimes u_{h}\otimes
\alpha_{0}\otimes \beta_{0} +\omega_{2} \sum_{h}gt_{h}\otimes
u_{h}\otimes \alpha_{2}\otimes \beta_{2} $$ $$ =\sum_{h}\omega_{0}
gt_{h}\otimes u_{h}\otimes \alpha_{0}\otimes \beta_{0}
+\sum_{h}\omega_{2}gt_{h}\otimes u_{h}\otimes \alpha_{2}\otimes
\beta_{2} $$ $$ =\sum_{h} (gt_{h}\otimes u_{h})\otimes (\omega_{0}
\alpha_{0}\otimes \beta_{0} +\omega_{2} \alpha_{2}\otimes
\beta_{2}) $$
\begin{equation}
=:\sum_{h} (gt_{h}\otimes u_{h})\otimes v'\label{2.9}
\end{equation}

The approximation issue is addressed by Theorem 2.1 below.

\begin{thm}
There is a constant $C>0$ so that for all unitary $g:
{\C}_{i}^{2}\otimes {\C}_{j}^{2} \rightarrow {\C}_{i}^{2}\otimes
{\C}_{j}^{2}$, there is a braid $b_{l}$ of length $\leq l$ in the
generators $\sigma_{i}$ and their inverses $\sigma_{i}^{-1}, 1\leq
i\leq n-1$, so that:
\begin{equation}
||\omega_{0} \rho^{0}(b_{l})-g\oplus id_{1}||+ || \omega_{2}
\rho^{2}(b_{l})-g\oplus id_{4}||\leq \epsilon (>0) \label{2.14}
\end{equation}
for some unit complex numbers (phases) $\omega_{i},i=0,2$ whenever
$\epsilon$ satisfies
\begin{equation}
l\leq C\cdot  \Big(\frac{1}{\epsilon}\Big)^{2}.\label{2.16}
\end{equation}
\end{thm}

We use $||\;\; ||$ to denote the operator norms and the subscripts
on $id$ indicate the dimension of the orthogonal component in
which we are trying {\it not} to act.

The main work in proving Theorem 2.1 is to show that the closure
of the image of the representation $\rho: B(6)\rightarrow
{\U}(5)\times {\U}(8)$ contains $SU(5)\times SU(8)$. Once this is
accomplished the estimate (\ref{2.14}) follows with some exponent
$\geq 2$ from [Ki] and the refinement to exponent=2 which will
appear in [CN] following a suggestion of the first author of the
present paper. Also as explained in [Ki] there is a
$poly(\frac{1}{\epsilon})$ time classical algorithm which can be
used to construct the approximating braid $b_{l}$ as a word in
$\{\sigma_{i}\}$ and $\{\sigma_{i}^{-1}\}$. The density theorem is
the substance of Section 4.

The action $\rho(b)$ $\lq\lq$approximately" executes the gate $g$
on $S_{k}$ but not in the usual sense of approximation since the
state space $i_{v}(S_{k})$ itself is only approximately $g$
invariant. To convert this $\lq\lq$smearing of qubits" to errors
of the type considered in the fault tolerant literature,  after
each $g$ is approximately executed by $\rho(b)$ we measure the
labels around $\cup_{i=1}^{k}\partial D_{i}$ to project the new
state $\rho(b)(s\otimes v)$ into the form $s'\otimes v, s'\in
S_{k}$, with probability $1-{\mathcal O}(\epsilon^{2}), |s'-s|\leq
{\mathcal O}( \epsilon)$. With probability ${\mathcal
O}(\epsilon^{2})$ the label measurement around $\partial D_{i}$
does not yield one;  in this case $V^{1}(D_{i}; \text{3
pts})=:V_{1,1,1,1}\cong \C^{2}$ has collapsed to $V_{1,1,1,3}\cong
\C$ and it is as if a qubit has been $\lq\lq$traced out" of our
state space. More specifically, if the label $3$ is measured on
$\partial D_{i}$, we replace $V^{3}(D_{i}, \text{its 3 marked
pts})$ with a freshly cooled qubit $V^{1}(D',\text{3 pts})$ with a
completely random initial state---an ancilli---which we have been
saving for such an occasion.  The reader may picture dragging
$D_{i}$ off to the edge of the disk $D$ and dragging the ancilli
$D_{i}'$ in as its replacement (and then renaming $D'$ by
$D_{i}$.)  The hypothesis that such ancilli are available is
discussed below.  The error model of [AB] is precisely suited to
this situation; Aharanov and Ben-Or show in Chapter 8 that a
calculation on the level of $\lq\lq$logical" qubits can be kept
precisely on track with a probability $\geq \frac{2}{3}$ provided
the ubiquitous errors at the level of $\lq\lq$physical" qubits are
of norm $\leq {\mathcal O}(\epsilon)$ (even if they are systematic
and not random) and the large errors (in our case tracing a qubit)
have probability also $\leq {\mathcal O}(\epsilon)$ for some
threshold constant $\epsilon >0$. For this, and all other fault
tolerant models, entropy must be kept at bay by ensuring a
$\lq\lq$cold" stream of ancilli $|0>$'s. In the context of our
model we must now explain both the role of measurement and
ancilla.

Given any essential simple closed curve $\gamma$ on a surface
$\Sigma$, the gluing formula reads:
\begin{equation}
 V(\Sigma)=\oplus_{l\in {\mathcal L}} V(\Sigma_{cut_{\gamma}},l)
\label{2.18}
\end{equation}
so $\lq\lq$measuring a label" means that we posit for every
$\gamma$ a Hermitian operator $H_{\gamma}$ with eigenvalues
distinguishing the summands of the r.h.s. of (\ref{2.18}) above.
For a more comprehensive computational study, we would wish to
posit that if $\gamma$ has length $=L$, then $H_{\gamma}$ can be
computed in poly(L) time.  For the present purpose we only need
that $H_{\gamma}, \gamma=\partial D_{i} \; \text{or} \;\partial
D_{i,j}$ can be computed in constant time.  Beyond measuring
labels, we hypothesize that there is some way of probing the
quantum state of the smallest nontrivial building blocks in the
theory.  For us these are the qubits $=V_{1,1,1,1}\cong {\C}^{2}$.
Fix a basis $\{|0>, |1>\}$ for $V_{1,1,1,1}$ and posit for each
$D_{i}, 1\leq i\leq k$, with label $1$ on its boundary, an
observable Hermitian operator $\sigma^{i}_{z}:
V_{3k}^{0}\rightarrow V_{3k}^{0}$ which acts as the Pauli matrix
 $\left( \begin{array}{cc} 1&0\\ 0& -1
\end{array} \right)$ in a fixed basis $\{ |0>, |1>\}$ for that
qubit.
  This is our repertoire of measurement:
$H_{\gamma}$ is used to $\lq\lq$unsmear physical qubits" after
each gate and the $\sigma_{z}$'s to read out the final state
(according to von Neumann's statistical postulate on measurement)
after the computation is completed.

In fault tolerant models of computation it is essential to have
available a stream of $\lq\lq$freshly cooled" ancilli qubits. If
these are present from the start of the computation, even if
untouched, they will decohere from errors in employing the
identity operator.  In the physical realization of a quantum
computer unless stored zeros were extremely stable there would
have to be some device (inherently not unitary!) for resetting
ancilli to $|0>$, e.g. a polarizing magnetic field.  As a
theoretical matter unbounded computation requires such resetting.
In a topological model such as $V(\Sigma)$ it is not unreasonable
to postulate that $|0>\in V_{1,1,1,1}=V^{1}(D_{i},\text{3 pts})$
is stable if not involved in any gates.  An alternative hypothesis
is that there is some mechanism outside the system analogous to
the polarizing magnetic field above which can $\lq\lq$refrigerate"
ancilli in the state $|0>$ until they are to be used. We refer
below to either of these as the $\lq\lq$fresh ancilli" hypothesis.
To correct the novel qubit smearing errors,  we already
encountered the need for ancilli in a random state $\rho=\left(
\begin{array}{cc} \frac{1}{2}&0\\ 0& \frac{1}{2}
\end{array} \right)$.  This state, of course, is easier to
maintain.

Let us now return to line (\ref{2.0}). Let $\U$ be the theoretical
output of a quantum circuit ${\mathcal C}$ of (i.e. composition of
) gates to be executed on the physical qubit level so as to
fault-tolerantly solve a problem instance of length $n$.  We
assume the problem is in $BQP$ and that the above composition has
length $\leq$ $poly(n)$.  Actually, due to error, $\mathcal C$
will output a completely positive trace preserving super-operator
$\mathcal O$, called a physical operator. Now simulate $\mathcal
C$ in the modular functor $V$ a gate at a time by a succession of
braidings and $H_{\gamma}$-measurements.  With regard to
parallelism (necessary in all fault tolerant schemes), notice that
disjoint 2 qubit gates can be performed simultaneously if
$D_{i,j}\cap D_{i',j'}=\emptyset$. For example this can always be
arranged in the linear QCM for gates acting in $D_{i, i+1}$ and
$D_{j, j+1}$ provided $i+1\neq j, j+1\neq i$, and $i\neq j$, and
even this model is shown to be fault tolerant [AB]. As noted
above,  the complementary vector $v\in V_{\text{complement}}$
evolves probabilistically as the simulation progresses . Different
$v$'s will occur as a tensor factor in a growing number of
probabilistically weighted terms. These new $v$-values are in the
end
 unimportant; they simply label a computational state
(to be observed with some probability) and are never read by the
output measurements $\sigma^{i}_{z}$.

Now the two main theorems:
\begin{thm}
 Let QCM denote the exact quantum circuit model.  Suppose $M$ is a
problem instance in BQP solved by a circuit $\mathcal C$ of length
poly(L) where L is length(M).  Let {\bf CS5} denote the model
based on the SU(2)-Chern-Simons modular functor of braids at the
fifth root of unity $e^{\frac{2\pi i}{5}}$ which we have described
in this section: uncolored 3k-strand braids, $H_{\gamma}$ and
$\sigma^{i}_{z}$ measurements, and $\lq\lq$fresh ancilli".  The
braid group acts on the modular functor and within the functor one
may identify k-qubits $S_{k}$.  These actions together with label
measurement $H_{\gamma}$'s define a probabilistic evolution of the
initial (possibly mixed) state $\alpha \in S_{k}$.  This
evolution, defined gate-wise, evolves the mixed state $\alpha
\otimes v\in V_{\text{3k}}$ of the modular functor to a new
(probabilistic mixture of ) state(s) $\beta$.  Performing
$\sigma^{i}_{z}$-measurements on $\beta$ samples from the mixture
drawing out a state $\beta_{l}=\alpha_{l}\otimes v_{l}$ and
observing (according to von Neumann measurement) only the
$\alpha_{l}$ factor.  With probability $\geq \frac{3}{4}$ the
observations correctly solve the problem instance $M$. The number
of marked points to be braided (=$\text{3k}$) and the length of
the braiding exceed the size of the original circuit $\mathcal C$
by at most a multiplicative poly(log(L)) factor. Taken in triples,
they represent the $\lq\lq$physical qubits" of the [AB] fault
tolerant model, thus {\bf CS5} provides a model which efficiently
and fault tolerantly simulates the computations of QCM.  We note
that the use of label measurements $H_{\gamma}$ introduces
non-unitary steps in the middle of our simulation.
\end{thm}

{\bf Proof:} The structure of the proof relies heavily on Chapter
8 [AB] to reduce the QCM to a linear quantum circuit (with state
space $S_{k}$) enjoying a very liberal error model (small
systematic errors plus rare trace over qubit).  In the final state
$\beta=\sum p_{l}\beta_{l}$, each $\beta_{l}$ admits a tensor
decomposition according to the geometry: $D=(\cup_{i}D_{i})\cup
(\text{complement})$, but along the k boundary components
$\cup_{i}\partial D_{i}$ all choices of labels 1 or 3 may appear.
So if we write $\beta_{l}=\alpha_{l}\otimes v_{l}$ we must
remember that associated to $l$ is an element $[l]\in \{1,3\}^{k}$
which defines the subspaces in which $\alpha_{l}$ and $v_{l}$ lie
and that $\beta_{l}$ lies in the corresponding $[l]$ sector of the
modular functor.  All occurrences of the label 3 correspond to a
$\C$ tensor factor, $\C\cong V^{3}(D_{i}, \text{3 pts}) \subset
V(D_{i}, \text{3 pts})$ whereas the label 1 corresponds to a
${\C}^{2}$ factor. Thus in the [AB] framework each label 3
corresponds to a $\lq\lq$lost" or averaged qubit according to our
replacement procedure $D_{i}\longleftrightarrow D'$. Losing an
occasional qubit from the computational space $S_{k}$ is the price
we pay to $\lq\lq$unsmear" $S_{k}$ within the modular functor.
Theorem 2.1 implies that for a braid length $=\mathcal
O(\frac{1}{\epsilon^{2}})$ a qubit will be lost with probability
$\mathcal O(\epsilon^{2})$ and if no qubit is lost the gate will
be performed with error $\mathcal O(\epsilon)$ on pure states.
Factoring a mixed state as a probabilistic combination of pure
states and passing the error estimate across the probabilities we
see that the $\mathcal O(\epsilon)$ error bound holds on the
super-operator trace norm as well. Thus for $\epsilon$
sufficiently
 small (estimated $< 10^{-6}$ in Chapter 8 [AB]),  observing
(at random) $\alpha_{l}$ amounts to sampling from an error prone
implementation of the quantum circuit $\mathcal C$.  The error
model is not entirely random in that the approximation procedure
used to construct $b_{L}$ will have systematic biases.  This
implies that the $\mathcal O(\epsilon)$ errors introduced in the
functioning of each gate are not random and must be treated as
$\lq\lq$malicious".  Fortunately the error model explained in
Chapter 8 [AB] permits the small error to be arbitrary as long as
the large error, e.g. qubit losses, occurs with a probability
dominated by a small constant independent of the qubit and the
computational history,  as they do in our {\bf CS5} model. This
completes the proof of Theorem 2.2 modulo the proof of the density
Theorem 4.1.

We may define a variant of our model $\bf CS5$, $\lq\lq$exact
Chern-Simons at $e^{\frac{2\pi i}{5}}$", $\bf ECS5$, in which we
assume that all the braid groups act exactly (no error) on the
modular functor $V$.  Such a hypothesis is not outrageous since a
physical implementation of a topological theory may itself confer
fault tolerance, in that topological phenomena are inherently
discrete.  The only difference in the algorithm for modeling the
QCM in $\bf ECS5$ is the simplification that $H_{\gamma}$
measurements are not performed in the middle of the simulation,
but only at the very end, prior to reading out the qubits $S_{k}$
with $\sigma_{z}^{k}$ measurements.

\begin{thm}
There is an efficient and strictly unitary simulation of QCM by
$\bf ECS5$.  Thus given a problem instance $M$ of length $L$ in
BQP, there is a classical poly(L) time algorithm for constructing
a braid $b$ as a word of length poly(L) in the generators
$\sigma_{i}, 1\leq i\leq \text{poly(L)=3k}$.  Applying $b$ to a
standard initial state, $\psi_{\text{initial}}\in
V^{0}(D,\text{3k})$, results in a state $\psi_{\text{final}}\in
V^{0}(D,\text{3k})$, so that the results of $H_{\gamma}$ on
$\partial D_{i}$ followed by $\sigma_{z}^{i}$ measurements on
$\psi_{\text{final}}$ correctly solve the problem instance $M$
with probability $\geq 6$.
\end{thm}

{\bf Proof:} In the quantum circuit model $\mathcal C$ for $M$
(implied by the problem lying in BQP) count the number $n$ of
gates to be applied.  Use line (11) to approximate each gate $g$
by a braid $b$ of length $l$ so that the operator norm error
$||\rho(b)-g||$ of the approximating gate will be less than
$\frac{1}{10n}$.  The composition of $n$ braids which gate-wise
simulate the quantum circuit introduces an error on operator norm
$<0.1$.  It follows that our two measurement steps will return an
answer (nearly) as reliable as the original quantum circuit
$\mathcal C$: $H_{\gamma}$ projects to $V^{1}(D, \text{3\; pts})$
with (more than) $90\%$ probability and the subsequent
probabilities of $\sigma_{z}^{1}$ measuring $|0>$ or $|1>$ differ
from $\mathcal C$ by less than $10\%$.

{\it Remark:} Theorem 2.2 and 2.3 are complementary.  One provided
additional fault tolerance---fault tolerance beyond what might be
inherent in a topological model---but at the cost of introducing
intermediate non-unitary steps (i.e. measurements).  The other
eschews intermediate measurements by and so gives a strictly
unitary simulation, but cannot confer additional fault tolerance.
It is an interesting open problem whether fault tolerance and
strictly unitary can be combined in a universal model of
computation based on topological modular functors.

\section{Jones' representation of the braid groups}

A TMF gives a family of representations of the braid groups and
mapping class groups. In this section, we identify the
representations of the braid groups from the SU(2) modular functor
at primitive roots of unity with the irreducible sectors of the
representation discovered by Jones whose weighted trace gives the
Jones polynomial of the closure link of the braid [J1][J2].  To
prove universality of the modular functor for quantum computation,
we only use this portion of the TMF.  Therefore, we will focus on
these representations.

First let us describe the Jones representation of the braid groups
explicitly following [We].  To do so, we need first to describe
the representation of the Temperley-Lieb-Jones algebras $A_{\beta,
n}$. Fix some integer $r\geq 3$ and $q=e^{\frac{2\pi i}{r}}$. Let
$[k]$ be the quantum integer defined as
$[k]=\frac{q^{\frac{k}{2}}-q^{\frac{-k}{2}}}{q^{\frac{1}{2}}-q^{\frac{-1}{2}}}$.
  Note that $[-k]=-[k]$, and $[2]=q^{\frac{1}{2}}+q^{\frac{-1}{2}}$.
Then $\beta :=[2]^2=q+\bar{q}+2=4cos^2(\frac{\pi}{r})$. The
algebras $A_{\beta, n}$ are the finite dimensional
${C}^{*}-$algebras generated by $1$ and projectors $e_1, \cdots ,
e_{n-1}$ such that
\begin{enumerate}
\item $e_i^2=e_i$, and $e_i^{*}=e_i$,

\item $e_i e_{i\pm 1} e_i=\beta ^{-1} e_i$,

\item $e_i e_j=e_j e_i$ if $|i-j|\geq 2$,

\end{enumerate}

and there exists a positive trace $tr: \cup_{n=1}^{\infty}
A_{\beta, n} \rightarrow \C$ such that $tr(xe_n)=\beta^{-1} tr
(x)$ for all $x\in A_{\beta, n}$.

The Jones representation of $A_{\beta, n}$ is the representation
corresponding to the G.N.S construction with respect to the above
trace. An important feature of the Jones representation is that it
splits as a direct sum of irreducible representations indexed by
some 2-row Young diagrams, which we will refer to as {\it
sectors}.  A Young diagram $\lambda=[\lambda_1, \cdots \lambda_s],
\lambda_{1}\geq \lambda_{2} \geq \cdots \geq \lambda_{s}$ is
called a $(2,r)$ diagram if $s\leq 2$ (at most two rows) and
$\lambda_1 -\lambda_2 \leq r-2$. Let $\wedge_n^{(2,r)}$ denote all
$(2,r)$ diagrams with $n$ nodes. Given $\lambda \in
\wedge_n^{(2,r)}$, let $T_{\lambda}^{(2,r)}$ be all standard
tableaus $\{t\}$ with shape $\lambda$ satisfying the inductive
condition which is the analogue of $(iii)$ in (3): when $n, n-1,
\cdots, 2,1$ are deleted from $t$ one at a time, each tableau
appeared is a tableau for some $(2,r)$ Young diagram. The
representation of $A_{\beta,n}$ is a direct sum of irreducible
representations $ \pi_{\lambda}^{(2,r)}$ over all $(2,r)$ Young
diagrams $\lambda$. The representation $\pi_{\lambda}^{(2,r)}$ for
a fixed $(2,r)$ Young diagram $\lambda$  is given as follows: let
$V_{\lambda}^{(2,r)}$ be the complex vector space with basis $\{
\vec{v}_{t}, \; t \in T_{\lambda}^{(2,r)}\}$. Given a generator
$e_{i}$ in the Temperley-Lieb-Jones algebra and a standard tableau
$t\in V_{\lambda}^{(2,r)}$. Suppose $i$ appears in $t$ in row
$r_1$ and column $c_1$, $i+1$ in row $r_2$ and column $c_2$.
Denote by $d_{t,i}=c_1 -c_2 -(r_1 -r_2)$,
$\alpha_{t,i}=\frac{[d_{t,i}+1]}{[2][d_{t,i}]}$, and
$\beta_{t,i}=\sqrt{\alpha_{t,i}(1-\alpha_{t,i})}$. They are both
non-negative real numbers and satisfy the equation
$\alpha_{t,i}=\alpha^2_{t,i} + \beta^2_{t,i}$. Then we define
\begin{equation}
 \pi_{\lambda}^{(2,r)}(e_{i}) (\vec{v}_t)=\alpha_{t,i}\vec{v}_t
+\beta_{t,i}\vec{v}_{g_{i}(t)}, \label{3.0}
\end{equation}
where $g_{i}(t)$ is the tableau obtained from $t$ by switching $i$
and $i+1$ if $g_{i}(t)$ is in $T_{\lambda}^{(2,r)}$.  If
$g_{i}(t)$ is not in $T_{\lambda}^{(2,r)}$, then $\alpha_{t,i}$ is
$0$ or $1$ given by its defining formula. This can occur in
several cases. It follows that $\pi_{\lambda}^{(2,r)}$ with
respect to the basis $\{\vec{v}_{t} \}$ is a matrix consisting of
only of $2\times 2$ and $1\times 1$ blocks. Furthermore, the $1
\times 1$ blocks are either $0$ or $1$, and the $2\times 2$ blocks
are
\begin{equation}
\left(
\begin{array}{cc}
\alpha_{t,i} & \beta_{t,i} \\
 \beta_{t,i}  & 1-\alpha_{t,i}
\end{array} \right). \label{3.1}
\end{equation}
The identity $\alpha_{t,i}=\alpha^2_{t,i} + \beta^2_{t,i}$ implies
that (\ref{3.1}) is a projector. So all eigenvalues of $e_{i}$ are
either $0$ or $1$.

The Jones representation of the braid groups is defined by
\begin{equation}
 \rho_{\beta,n}(\sigma_i)=q-(1+q) e_i. \label{3.2}
\end{equation}
Combining (\ref{3.2}) with the above representation of the
Temperley-Lieb-Jones algebra, we get Jones' representation of the
braid groups, denoted still by $\rho_{\beta, n}$: $$\rho_{\beta,
n} : B_n \rightarrow A_{\beta, n} \rightarrow \U(N_{\beta, n}),$$
 where the dimension $N_{\beta,n}=\sum_{\lambda \in \wedge_{n}^{(2,r)}}
\text{dim} V_{\lambda}^{(2,r)}$ grows asymptotically as
$\beta^{n}$.

 When $|q|=1$, as we have seen already, Jones' representation
 $\rho_{\beta, n}$ is
unitary.  To verify that $\rho(\sigma_{i})\rho^{*}(\sigma_{i})=1$,
note $\rho^{*}(\sigma_{i})=\bar{q}-(1+\bar{q})e^{*}_{i}$. So we
have $\rho(\sigma_{i})\rho^{*}(\sigma_{i})=
q\bar{q}+(1+q)(1+\bar{q})e_{i}e_{i}^{*}-(1+q)e_{i}-(1+\bar{q})e_{i}^{*}=
1$.  We use the fact $e_{i}^{*}=e_{i}$ and $e_{i}^{2}=e_{i}$ to
cancel out the last 3 terms.

From the definition, $\rho_{\beta, n}$ also splits as a direct sum
of representations over $(2,r)$-Young diagrams. A sector
corresponding to a particular Young diagram $\lambda$ will be
denoted by $\rho_{\lambda,\beta, n}$.

Now we collect some properties about the Jones representation of
the braid groups into the following:
\begin{thm}
(i) For each $(2,r)$-Young diagram $\lambda$, the representation
$\rho_{\lambda,\beta, n}$ is irreducible.

(ii) The matrices $\rho_{\lambda,\beta,  n}(\sigma_i)$ for $i=1,2$
generate an infinite subgroup of $\U (2)$ modulo center for $r\neq
3,4,6,10$.

(iii) Each matrix $\rho_{\lambda,\beta, n}(\sigma_i), 1\leq i\leq
n-1,$ has exactly two distinct eigenvalues $-1, q$.

(iv)  For the (2,5)-Young diagram $\lambda=[4,2]$, $n=6$, the two
eigenvalues $-1,q$ of every $\rho_{\lambda,\beta,  6}(\sigma_i)$
have multiplicity of 3 and 5, respectively.

\end{thm}

The proofs of (i) and (ii) are in [J2]. For (iii), first note that
the matrix $\rho_{\lambda, \beta, n}(\sigma_{1})$ is a diagonal
matrix with respect to the basis $\{\vec{v}_{t}\}$ with only two
distinct eigenvalues $-1,q$.  Now (iii) follows from the fact that
all braid generators $\sigma_{i}$ are conjugate to each other. For
(iv), simply check the explicit matrix for $\rho_{\lambda,\beta,
6}(\sigma_1)$ at the end of this section.

Now we identify the sectors of the Jones  representation with the
representations of the braid groups coming from the $SU(2)$
Chern-Simons modular functor. The $SU(2)$ Chern-Simons modular
functor $\bf CSr$ of level $r$ has been constructed several times
in the literature (for example, [RT][T][Wa][G]).
Our construction of the modular functor $\bf CSr$ is based on
skein theory [KL].  The key ingredient is the substitute of
Jones-Wenzl idempotents for the intertwiners of the irreducible
representations of quantum groups [RT][T][Wa]. This is the same
$SU(2)$ modular functor as constructed using quantum groups in
[RT] (see [T]) which is regarded as a mathematical realization of
the Witten-Chern-Simons theory.  All formulae we need for skein
theory are summarized in Chapter 9 of [KL] with appropriate
admissible conditions. Fix an integer $r\geq 3$.  Let
$A=\sqrt{-1}\cdot e^{-\frac{2\pi i}{4r}}$, and $s=A^{2}$, and
$q=A^{4}$. (Note the confusion caused by notations. The $q$ in
[KL] is $A^{2}$ which is our $s$ here. But in Jones'
representation of the braid groups [J2], $q$ is $A^{4}$. In all
formulae in [KL], $q$ should be interpreted  as $s$ in our
notation.)
  The label set $\mathcal L$ of the
modular functor $\bf CSr$ will be $\{0,1,\cdots, r-2\}$ and the
involution is the identity.  We are interested in a unitary
modular functor and the one in [G] is not unitary.  We claim that
if we follow the same construction of [G] using our choice of $A$
and endow all state spaces of the modular functor with the
following Hermitian inner product, the resulting modular functor
$\bf CSr$ is unitary.

Given a surface $\Sigma$,  a pants decomposition of $\Sigma$
determines a basis of $V(\Sigma)$: each basis element is a tensor
product of the basis elements of the constituent pants. The
desired inner products are determined by axiom $(2.14)$ [Wa] if we
specify an inner product on each space $V_{abc}$.  Our choice of
$A$ makes all constants $S(a)$ appearing in the axiom $(2.14)$
[Wa] positive. Consequently, positive definite Hermitian inner
products on all spaces $V_{abc}$ determine a positive definite
Hermitian inner product on $V(\Sigma)$.  The vector space
$V_{abc}$ of the three punctured sphere $P_{abc}$ is defined to be
the skein space of the disk $D_{abc}$ enclosed by the seams of the
punctured sphere $P_{abc}$. The numbering of the three punctures
induces a numbering of the three boundary $\lq\lq$points" of the
disk $D_{abc}$ labeled by $\{a,b,c\}$. Suppose $t$ is a tangle on
$D_{abc}$ in the skein space of $D_{abc}$, and let $\bar{t}$ be
the tangle on $D_{abc}$ obtained by reflecting the disk $D_{abc}$
through the first boundary point and the origin. Then the inner
product $<,>_{h}: V_{abc}\times V_{abc} \rightarrow \C$ is as
follows: given two tangles $s$ and $t$ on $D_{abc}$, their product
$<s,t>_{h}$ is the Kauffman bracket evaluation of the resulting
diagram on $S^{2}$ obtained by gluing the two disks with $s$ and
$\bar{t}$ on them respectively, along their common boundaries with
matching numberings. Extending $<,>_{h}$ on the skein space of
$D_{abc}$ linearly in the first coordinate and conjugate linearly
in the second coordinate, we obtain a positive definite Hermitian
inner product on $V_{abc}$.  It is also true that the mapping
class groupoid actions in the basic data respect this Hermitian
product, and the fusion and scattering matrices $F$ and $S$ also
preserve this product.  So $\bf CSr$ is indeed a unitary modular
functor.

This modular functor $\bf CSr$ defines representations of the
central extension of the mapping class groups of labeled extended
surfaces, in particular for $n$-punctured disks $D_{n}^{m}$ with
all interior punctures labeled 1 and boundary labeled $m$. If
$m\neq 1$, then the mapping class group is the braid group
$B_{n}$. If $m=1$, then the mapping class group is the spherical
braid group ${\mathcal S}B_{n+1}=\mathcal M(0,n+1)$. Recall that
we suppress the issues of framing and central extension as they
are inessential in our discussion. Also the representation of the
mapping class groups coming from $\bf CSr$ will be denoted simply
by $\rho_{r}$.

\begin{thm}
Let $D_{n}^{m}$ be as above.

(1): If $m+n$ is even, and $m\neq 1$, then $\rho_{r}$ is
equivalent to the irreducible sector of the Jones representation
$\rho_{\lambda, \beta, n}$ for the Young diagram
$\lambda=[\frac{m+n}{2}, \frac{m-n}{2}]$ up to phase.

(2): If $n$ is odd, and $m=1$, then the composition of $\rho_{r}$
with the natural map $\iota:B_{n}\rightarrow {\mathcal S}B_{n+1}$
is equivalent to the irreducible sector of the Jones
representation $\rho_{\lambda, \beta, n}$ for the Young diagram
$\lambda=[\frac{n+1}{2}, \frac{n-1}{2}]$ up to phase.
\end{thm}

The equivalence of these two representations was first established
in a non-unitary version [Fu]. A computational proof of this
theorem can be obtained following [Fu]. So we will be content with
giving some examples for $r=5$.

For the $(2,5)$ Young diagram $\lambda=[2,1]$, $n=3$ with an
appropriate ordering of the basis:

$\rho_{[2,1], \beta, 3}(\sigma_{1})=\left(
\begin{array}{cc}
-1 & 0\\ 0& q
\end{array} \right),
$

$\rho_{[2,1], \beta, 3}(\sigma_{2})=\left(
\begin{array}{cc}
\frac{q^2}{q+1} & -\frac{q\sqrt{[3]}}{q+1}\\
-\frac{q\sqrt{[3]}}{q+1}&-\frac{1}{q+1}
\end{array} \right),
$
where quantum $[3]=q+\bar{q}+1.$

For the $(2,5)$ Young diagram $\lambda=[3,3]$, $n=6$, the
representation is 5-dimensional.  With an appropriate ordering of
the basis, we have:

$\rho_{[3,3], \beta, 6}(\sigma_{1})=\left(
\begin{array}{ccccc}
-1 &  & &&\\ & q &&&\\ && -1&&\\ &&& q&\\ &&&& q
\end{array} \right),
$

$\rho_{[3,3], \beta, 6}(\sigma_{2})=\left(
\begin{array}{ccccc}
\frac{q^2}{q+1} & -\frac{q\sqrt{[3]}}{q+1} &&& \\
-\frac{q\sqrt{[3]}}{q+1}&-\frac{1}{q+1} &&& \\ && \frac{q^2}{q+1}
& -\frac{q\sqrt{[3]}}{q+1} & \\
&&-\frac{q\sqrt{[3]}}{q+1}&-\frac{1}{q+1} & \\ &&&& q
\end{array} \right).
$

For the $(2,5)$ Young diagram $\lambda=[4,2]$, $n=6$, the
representation is 8-dimensional.  Here the inductive condition on
basis elements make one standard tableau illegal, so the
representation is not 9-dimensional as it would be if $r>5$.  This
is the restriction analogous to $(iii)$ in (3) for the modular
functor. With an appropriate ordering of the basis:

$\rho_{[4,2], \beta, 6}(\sigma_{1})=\left(
\begin{array}{cccccccc}
-1 &  & && &&&\\
 & q &&& &&&\\
 &&-1 &&&&&\\
 && &q&&&&\\
  &&&& -1&&&\\
  &&&&&q &&\\
  &&&&&&q&\\
   &&&&&&& q
\end{array} \right).
$

\section{A Density theorem}

In this section, we prove the density theorem.
\begin{thm}
Let $\rho:=\rho_{[3,3]}\oplus \rho_{[4,2]}: B_{6}\rightarrow
\U(5)\times \U(8)$ be the Jones representation of $B_{6}$
 at the $5$-th root of unity $q=e^{\frac{2\pi i}{5}}$.
Then the closure of the image of $\rho(B_{6})$ in $\U(5)\times
\U(8)$ contains $SU(5)\times SU(8)$.
\end{thm}

By Theorem 3.2, this is the same representation
 $\rho:=\rho^{0}\oplus \rho^{2}: B_{6}\rightarrow
\U(5)\times \U(8)$  in the $SU(2)$ Chern-Simons modular functor at
the $5$-th root of unity used in Section 2 to build a universal
quantum computer.  In the following, a key fact used is that the
image matrix of each braid generator under the Jones
representation has exactly two eigenvalues $\{-1, q\}$ whose ratio
is not $\pm 1$.  This strong restriction allows us to identify
both the closed image and its representation.

{\bf Proof:}  First it suffices to show that the images of
$\rho_{[3,3]}$ and $\rho_{[4,2]} $ contain $SU(5)$ and $SU(8)$,
respectively.  Supposing so, if $K=\overline{\rho(B_{6})}\cap
(SU(5)\times SU(8))$, then the two projections
 $p_{1}:K\rightarrow SU(5)$ and
 $p_{2}:K\rightarrow SU(8)$ are both surjective.
Let $N_{2}$ (respectively $N_{1}$) be the kernel of $p_{1}$
(respectively $p_{2}$). Then $N_{1}$ (respectively $N_{2}$) can be
identified as a normal subgroup of $SU(5)$ (respectively $SU(8)$).
By Goursat's Lemma (page 54, [La]), the image of $K$ in
$SU(5)/N_{1}\times SU(8)/N_{2}$ is the graph of some isomorphism
$SU(5)/N_{1} \cong SU(8)/N_{2}$. As the only nontrivial normal
subgroups of $SU(n)$ are finite groups, this is possible only if
$N_{1}=SU(5)$ and $N_{2}=SU(8)$. Therefore, $K=SU(5)\times SU(8)$.

The proofs of the density for $\rho_{[3,3]}$ and $\rho_{[4,2]}$
are similar.  So we prove both cases at the same time and give
separate argument for the more complicated case $\rho_{[4,2]}$
when necessary.

Let $G$ be the closure of the image of $\rho_{[3,3]}$ (or
$\rho_{[4,2]}$) in $\U(5)$ (or $\U(8)$) which we will try to
identify. By Theorem 3.1, G is a compact subgroup of $\U(m)\;(
m=5\; \text{or}\;8)$ of positive dimension. Denote by $V$ the
induced $m$-dimensional faithful, irreducible complex
representation of $G$. The representation $V$ is faithful since
$G$ is a subgroup of $\U(m)$. Let $H$ be the identity component of
$G$. What we actually show is that the derived group of $H$,
$Der(H)=[H,H]$, is actually $SU(m)$. We will divide the proof into
several steps.

{\bf Claim 1:} The restriction of $V$ to $H$ is an isotypic
representation, i.e. a direct sum of several copies of a single
irreducible representation of $H$.

{\bf Proof:} As $G$ is compact,  $V=\oplus_{P}V_{P}$, where $P$
runs through some irreducible representations of $H$, and $V_P$ is
the direct sum of all the copies of $P$ contained in $V$. Since
$H$ is a normal subgroup, and the braid generators $\sigma_i$
topologically generate $G$, the $\sigma_i$'s permute transitively
the isotypic components $V_P$ [CR, Section 49]. If there is more
than 1 such component, then some $\sigma_i$ acts nontrivially, so
it must permute these blocks.

Now we need a linear algebra lemma:
\begin{lemm}
Suppose $W$ is a vector space with a direct sum decomposition
$W=\oplus_{i=1}^{n} W_{i}$, and there is a linear automorphism $T$
such that $T:W_{i}\rightarrow W_{i+1}$ $1\leq i\leq n$ cyclically.
Then the product of any eigenvalue of $T$ with any $n$-th root of
unity is still an eigenvalue of $T$.
\end{lemm}

{\bf Proof:} Choose a basis of $W$ consisting of bases of
$W_{i},i=1,2,\cdots, n$. If $k$ is not a multiple of $n$, then
$\text{tr}T^k=0$, as all diagonal entries are 0 with respect to
the above basis.  Let $\{\lambda_{i}\}$ be all eigenvalues of $T$.
( They may repeat.)  Consider all values of $\text{tr}T^m=\sum
{\lambda_{i}}^{m}\; (m=1,2,\cdots) $ which are sums of $m$-th
powers of all eigenvalues of $T$.  These sums of $m$-th powers of
$\{\lambda_{i}\}$  are invariant if we simultaneously multiply all
the eigenvalues $\{ \lambda_{i} \}$ by an $n$-th root of unity
$\omega$:  $\sum {(\omega \lambda_{i})}^{m}=\sum\omega^{m}
{\lambda_{i}}^{m}={\omega}^{m}\sum {\lambda_{i}}^{m}$ which is
equal to $\text{tr}T^m=\sum {\lambda_{i}}^{m}$ because when $m$ is
not a multiple of $n$, they are both $0$, and when $m$ is,
$\omega^{m}=1$. These values $\text{tr}T^{m}$ uniquely determine
the eigenvalues of $T$, and therefore the set of the eigenvalues
of $T$ is invariant under multiplication by any $n$-th root of
unity.

Back to claim 1, if there is more than one isotypic component,
then some $\sigma_{i}$ will have an orbit of length at least 2. It
is impossible to have an orbit of length 3 or more by the above
lemma as this will lead to at least 3 eigenvalues.  If the orbit
is of length 2 and as $\rho(\sigma_{i})$ has only two eigenvalues
$\{a,b\}$, by the lemma, $\{-a,-b\}$ are also eigenvalues.  It
follows that $a=-b$ which is impossible when $q\neq -1$.

{\bf Claim 2:} The restriction of $V$ to $H$ is an irreducible
representation.

{\bf Proof:}  By claim 1, $V|_{H}$ has only one isotypic
component. If $V|_{H}$ is reducible, then the isotypic component
is a tensor product $V_{1}\otimes V_{2}$, where $V_{1}$ is the
irreducible representation of $H$ in the isotypic component and
$V_{2}$ is a trivial representation of $H$ with
$\text{dim}V_{2}\geq 2$. If $V_{1}$ is 1-dimensional, then
$\rho(\sigma_{i}), i=1,2$ generate a finite subgroup of $\U(m)$
modulo center which is excluded by Theorem 3.1.  So we have
$\text{dim}V_{1} \geq 2$. Now we recall a fact in representation
theory: a representation of a group $\rho: G\rightarrow GL(V)$ is
irreducible if and only if the image $\rho(G)$ of $G$ generates
the full matrix algebra $\text{End}(V)$. As $V_{1}$ is an
irreducible representation of $H$,  the image $\rho(H)$ generates
$\text{End}(V_{1})\otimes \text{id}_{2}$, where the subscript of
$id$ indicate the tensor factor. As the elements $\sigma_{i}$
normalize $H$, they also normalize the subalgebra
$\text{End}(V_{1})\otimes \text{id}_{2}$ in
$\text{End}(V_{1}\otimes V_{2})$.  Consequently they act as
automorphisms of the full matrix algebra $\text{End}(V_{1})$. Any
automorphism of a full matrix algebra is a conjugation  by a
matrix, so the braid generators $\sigma_{i}$ act via conjugation
(up to a scalar multiple) as invertible matrices in
$\text{End}(V_{1})\otimes \text{id}_{2}$
 modulo its centralizer.
It is not hard to see the centralizer of $\text{End}(V_{1})\otimes
\text{id}_{2}$ in $\text{End}(V_{1}\otimes V_{2})$ is
$\text{id}_{1} \otimes \text{End}(V_{2})$.  Therefore, the braid
generators $\sigma_{i}$ act via conjugation as invertible matrices
in $\text{End}(V_{1})\otimes \text{End}(V_{2})$, i.e. they
preserve the tensor decomposition.  This is impossible by the
following eigenvalue analysis.  Consider a braid generator
$\sigma_{i}$, its image $\rho(\sigma_{i})$ is a tensor product of
two matrices each of sizes at least 2. Since $\rho(\sigma_{i})$
has only two eigenvalues, neither factor matrix can have 3 or more
eigenvalues. If both factor matrices have two eigenvalues, the
fact that $\rho(\sigma_{i})$ has 2 eigenvalues in all implies that
the ratio of these two eigenvalues is $\pm 1$ which is forbidden.
If one factor matrix is trivial, then $\rho(\sigma_{i})$ acts
trivially on this factor. As all braid generators are conjugate to
each other, so the whole group $G$ will act trivially on this
factor which implies that $V$ is a reducible representation of
$G$. This case cannot happen either, as $V$ is an irreducible
representation of $G$.

{\bf Claim 3:} The derived group, $Der(H)=[H,H]$,  of $H$ is a
semi-simple Lie group, and the further restriction of $V$ to
$Der(H)$ is still irreducible.

{\bf Proof:} By claim 2,  $V|_{H}$ is a faithful, irreducible
representation, so $H$ is a reductive Lie group [V, Theorem
3.16.3]. It follows that the derived group of $H$ is semi-simple.
It also follows that the derived group and the center of $H$
generate $H$. By Schur's lemma, the center act by scalars. So
$V|_{Der(H)}$ is still irreducible.

{\bf Claim 4:}  Every outer automorphism of $Der(H)$ has order 1,
2, or 3.

First we recall a simple fact in representation theory.  If $V$ is
an irreducible representation of a product group $G_{1}\times
G_{2}$, then $V$ splits as an outer tensor product of irreducible
representations of $G_{i}, i=1,2$. The restriction of $V$ to
$G_{1}$ has only one isotypic component, and the restriction of
$V$ to $G_{2}$ lies in the centralizer of the image of $G_{1}$. So
the representation splits.

{\bf Proof:} It suffices to prove the same statement for the
universal covering $Der^{uc}(H)$ of $Der(H)$, as the automorphism
group of $Der(H)$ is a subgroup of the automorphism group of
$Der^{uc}(H)$.

For the 5-dimensional case: as 5 is a prime, $Der^{uc}(H)$ is a
simple group.  It is well-known that any outer automorphism of a
simple Lie group is of order 1, 2, or 3.

For the 8-dimensional case, if $Der^{uc}(H)$ is a simple group, it
can be handled as above, so we need only to consider the split
cases. If $Der^{uc}(H)$ splits into two simple factors, then one
factor must be $SU(2)$: of all simply connected simple Lie groups,
only $SU(2)$ has a 2-dimensional irreducible representation. So
the outer automorphism group is either $Z_{2}$ when both factors
are $SU(2)$, or the same as the outer automorphism group of the
other simple factor. Our claim holds. If there are three simple
factors, they must all be $SU(2)$. The outer automorphism group is
the permutation group on three letters $S_{3}$.  Again our claim
is true.

{\bf Claim 5:} For each braid generator $\sigma_{i}$, we can
choose a corresponding element $\tilde{\sigma_{i}}$ lying in the
derived group $Der(H)$ which also has exactly two eigenvalues,
whose ratio is not $\pm 1$.  The multiplicity of each eigenvalue
of $\tilde{\sigma_{i}}$ is the same as that of $\sigma_{i}$. (The
choice of $\tilde{\sigma_{i}}$ is not unique, but its two
eigenvalues have ratio $q$.)

{\bf Proof:}  Since $Der(H)$ is still a normal subgroup of $G$,
and
 the braid generators $\sigma_i$
normalize $Der(H)$, so they determine outer-automorphisms of
$Der(H)$. By claim 4, an outer-automorphism of $Der(H)$
 is of order 1, 2, or 3.  Hence
$\sigma^{6}_{i}$ acts as an inner automorphism of $Der(H)$. By
Schur's lemma, each $\sigma^{6}_i$ is the product of an element in
$Der(H)$ with a scalar, though the decomposition is not unique.
Fix a choice for an element $\tilde{\sigma_{i}}$ in $Der(H)$. Then
it has exactly two desired eigenvalues.

To complete the proof of Theorem 4.1, we summarize our situation:
we have a nontrivial semi-simple group $Der^{uc}(H)$ with an
irreducible unitary representation. Furthermore, it has a special
element $x$ whose image under the representation has exactly two
distinct eigenvalues whose ratio is not $\pm 1$.

For the 5-dimensional case, $Der^{uc}(H)$ is a simple Lie group.
Going through the list [MP] of pairs $(G,\varpi)$, where $G$ is a
simply connected Lie group and $\varpi$ a dominant weight. The
only possible 5-dimensional irreducible representations are as
follows: rank=1, $(SU(2), {4}\varpi_{1})$, rank=2, $(Sp(4),
\varpi_{2})$, and rank=4, $(SU(5), \varpi_{i}), i=1,4$. By
examining the possible eigenvalues, we can exclude the first two
cases as follows: for the first case, suppose $\alpha, \beta$ are
the two eigenvalues of the above element $x$ in $SU(2)$, then
under the representation ${4}\varpi_{1}$ the eigenvalues of the
image of $x$ are $\alpha^i \beta^j, i+j=4$, where $i$ and $j$ both
are non-negative integers. The only possibility is two eigenvalues
whose ratio is $\pm 1$. For the second case, since $5$ is an odd
number, any element in the image has a real eigenvalue. Other
eigenvalues come in mutually reciprocal pairs.  Again the only
possibility is two eigenvalues whose ratio is $\pm 1$. Therefore,
the only possible pair is the third case which gives
$Der^{uc}(H)=SU(5)$. As $V$ is a faithful representation of
$Der(H)$, the image of $Der(H)$ is the same as that of
$Der^{uc}(H)$ which is $SU(5)$.

The 8-dimensional case for $\rho_{[4,2]}$ is similar.  By [MP], we
see the possible pairs for simply connected simple groups are
$\big( SU(2), 7{\varpi_{1}}\big)$, $\big(SU(3), \varpi_{1}+\varpi_{2}\big)$,
$\big(Spin(7), \varpi_{3}\big)$, $\big(Sp(8), \varpi_{1}\big)$, $\big(Spin(8),
\varpi_{i}\big), i=1,3,4$ and $\big(SU(8), \varpi_{i}\big),i=1,7$, where
$\varpi_{i}$ is the fundamental weight. The same eigenvalue
analysis will exclude all but the $\big(SU(8), \varpi_{i}\big)$ case. The
proof follows the same pattern as above with the following
novelties. Case 2 is the adjoint representation of $SU(3)$, if the
special element $x\in SU(3)$ has eigenvalues $\{\alpha, \beta,
\gamma \}$,  the image matrix of $x$ will have eigenvalue $1$ with
multiplicity 2 and all six pair-wise ratios of
$\{\alpha,\beta,\gamma\}$, so they are $\pm 1$. For case 4, recall
that if $\lambda$ is an eigenvalue of a symplectic matrix, so is
$\lambda^{-1}$ with the same multiplicity, thus there are
candidates for the special element $x$, but all such elements have
the property that the multiplicity for both eigenvalues is 4.
Notice by Theorem 3.1 (iv), the multiplicity of the two distinct
eigenvalue in $\tilde{\rho}(\sigma_{i})$ is 3 and 5, respectively.
Case 5 is done just as case 4. This excludes all the unwanted
simple groups. We have to consider also the product cases. For
product of two or three simple factors, the same analysis of
eigenvalues as at the end of the proof of claim 2 excludes them.
Actually, there are only four cases here: $SU(2)\times SU(2)$,
$SU(2)\times SU(4)$, $SU(2)\times Sp(4)$ and $SU(2)\times
SU(2)\times SU(2)$. This completes the proof of our density
theorem.


\begin{thebibliography} {[BK]}

\bibitem[AB]{AB}D. Aharanov and M. Ben-Or, {\it Fault tolerant
quantum computation with constant error}, quant-ph/9906129.
\bibitem[CN]{CN}I. Chuang and M. Nielsen, Quantum computation and Quantum
information,  Cambridge Univ. Press (a book to appear).
\bibitem[CR]{CR}C. Curtis and I. Reiner, {\it Representation theory of
finite groups and associate algebras},  Pure and Applied Math.,
vol XI, Interscience Publisher, 1962.
\bibitem[D]{D}D. Deutsch, {\it Quantum computational networks,}
Proc. Roy. Soc. London, {\bf A425}(1989), 73-90.
\bibitem[Fey]{Fey}R. Feynman, {\it Simulating physics with computers,}
Int. J. Theor. Phys. {\bf 21}(1982), 467-488.
\bibitem[FKW]{FKW}M. Freedman, A. Kitaev, and Z. Wang, {\it
Simulation of topological field theories by quantum computers},
quant-ph/0001071.
\bibitem[Fu]{Fu}L. Funar, {\it On the TQFT representations of the
mapping class groups}, Pac. J. Math., {\bf 188}(1999), 251-274.
\bibitem[G]{G}R. Gelca, {\it Topological quantum field theory with
corners based on the Kauffmann bracket}, Comment. Math. Helv.,
{\bf 72}(1997), 210-243.
\bibitem[J1]{J1} V.F.R. Jones, {\it Hecke algebra representations
of braid groups and link polynomial,} Ann. Math., {\bf 126}(1987),
335-388.
\bibitem[J2]{J2}V.F.R. Jones, {\it Braid groups, Hecke algebras and
type $II_1$ factors}, Geometric methods in operator algebras,
Proc. of the US-Japan Seminar, Kyoto, July 1983.
\bibitem[KL]{KL}L. Kauffmann and S. Lins, {\it Temperley-Lieb recoupling
theory and invariants of 3-manifolds}, Ann. Math. Studies, vol
134, Princeton Univ. Press, 1994.
\bibitem[Ki]{Ki}A. Kitaev,  {\it Quantum computations: algorithms
and error correction,} Russian Math. Survey, {\bf 52:61}(1997),
1191-1249.
\bibitem[La]{La}S. Lang, {\it Algebra}, 2nd edition, Addison-Wesley
Publishing Company, 1984.
\bibitem[Ll]{Ll}S. Lloyd, {\it Universal quantum simulators,}
Science, {\bf 273}(1996), 1073-1078.
\bibitem[MP]{MP}W. Mckay and J. Patera, {\it Tables of dimensions,
indices, and branching rules for representations of simple Lie
algebras}, Lecture Notes in Pure and Applied Math., vol 69.
\bibitem[RT]{RT}N. Reshetikhin, and V.G. Turaev,
{\it Invariants of $3$-manifolds via link polynomials and quantum
groups}, Invent. Math. {\bf 103} (1991), no. 3, 547--597.
\bibitem[T]{T}V. Turaev, {\it Quantum invariants of knots and
3-manifolds}, de Gruyter Studies in Math., vol 18, 1994.
\bibitem[V]{V}V.S. Varadarajan,{\it Lie groups, Lie algebras and
their representaions}, Graduate Texts in Math., vol 102,
Springer-Verlag, 1984.
\bibitem[Wa]{Wa}K. Walker, {\it On Witten's 3-manifold invariants,}
preprint, 1991.
\bibitem[We]{We}H. Wenzl, {\it Hecke algebras of type $A_n$ and
subfactors}, Invent. Math. {\bf 92}(1988), 349-383.
\bibitem[Wi]{Wi}E. Witten, {\it Quantum field theory and the Jones
polynomial,} Comm. Math. Phys., {\bf 121}(1989), 351-399.
\bibitem[Y]{Y}A. Yao, {\it Quantum circuit complexity,} Proc. 34th
Annual Symposium on Foundations of Computer Science, IEEE Computer
Society Press, Los Alamitos, CA, 352-361.

\end{thebibliography}
\end{document}